\documentclass[aps,prb,reprint,twocolumn,superscriptaddress,showpacs,nofootinbib]{revtex4-2}
\usepackage{amsmath,amssymb}
\usepackage{makecell}
\usepackage{multirow}
\usepackage{CJK}
\usepackage{graphicx}
\usepackage{mathrsfs}
\usepackage{bm}
\usepackage{amsmath}
\usepackage{dcolumn}
\usepackage{epstopdf}
\usepackage{dsfont}
\usepackage{amssymb}
\usepackage{tabularx}
\usepackage{array}
\usepackage{float}
\usepackage{color}
\usepackage{epstopdf}
\usepackage{mathrsfs}
\usepackage[colorlinks, linkcolor=blue,anchorcolor=blue,citecolor=blue,urlcolor=blue]{hyperref}
\usepackage{extarrows}
\usepackage{braket}

\begin{document}
\title{Quadratic nodal point in a two-dimensional noncollinear antiferromagnet}
\author{Xukun Feng}
\affiliation{Research Laboratory for Quantum Materials, Singapore University of Technology and Design, Singapore 487372, Singapore}

\author{Zeying Zhang}
\email{zzy@mail.buct.edu.cn}
\affiliation{College of Mathematics and Physics, Beijing University of Chemical Technology, Beijing 100029 , China}
\affiliation{Research Laboratory for Quantum Materials, Singapore University of Technology and Design, Singapore 487372, Singapore}

\author{Weikang Wu}
\email{weikang\_wu@sdu.edu.cn}
\affiliation{Key Laboratory for Liquid-Solid Structural Evolution and Processing of Materials
(Ministry of Education), Shandong University, Jinan 250061, China}

\author{Xian-Lei Sheng}
\affiliation{School of Physics, Beihang University, Beijing 100191, China}
\affiliation{Peng Huanwu Collaborative Center for Research and Education, Beihang University, Beijing 100191, China}

\author{Shengyuan A. Yang}
\affiliation{Research Laboratory for Quantum Materials, IAPME, University of Macau, Macau SAR}
\affiliation{Research Laboratory for Quantum Materials, Singapore University of Technology and Design, Singapore 487372, Singapore}

\begin{abstract}
Quadratic nodal point (QNP) in two dimensions has so far been reported only in nonmagnetic materials and in the absence of spin-orbit coupling. Here, by first-principles calculations and symmetry analysis, we predict stable QNP near Fermi level in a two-dimensional kagome metal-organic framework material, Cr$_3$(HAB)$_2$, which features noncollinear antiferromagnetic ordering and sizable spin-orbit coupling. Effective $k\cdot p$ and lattice models are constructed to capture such magnetic QNPs.
Besides QNP, we find Cr$_3$(HAB)$_2$ also hosts six magnetic linear nodal points protected by mirror as well as $C_{2z}T$ symmetry.
Properties associated to these nodal points, such as topological edge states and quantized optical absorbance, are discussed.


\end{abstract}
\maketitle

\section{Introduction}

In the study of topological semimetals, the focus is on band degeneracies near the Fermi level, around which the electronic states can acquire multi-component pseudospin structures and may exhibit unusual physical properties~\cite{chiu2016classification,dai2016weyl,bansil2016colloquium,armitage2018weyl,lv2021experimental}.
For example, in Weyl semimetals, the bands at Fermi level form doubly degenerate Weyl nodal points, such that the low-energy electrons resemble Weyl fermions from high-energy physics~\cite{wan2011topological,murakami2007phase}. In most cases, the formation of band degeneracies requires protection by crystalline symmetries~\cite{chiu2016classification}. For example, a stable band degeneracy at a high-symmetry point of Brillouin zone (BZ) corresponds to an irreducible representation for the little co-group at that point. It follows that the allowed types of band degeneracies must strongly depend on factors such as dimensionality of system, existence of magnetism, and spin-orbit coupling (SOC), since they affect the symmetry groups and their representations~\cite{yu2022encyclopedia,liu2022systematic,zhang2022encyclopedia,tang2022complete,knoll2022classification}.

The band dispersion around a nodal point is typically of linear type. However, it was found that some special crystal symmetries may protect nodal points with higher-order dispersions \cite{xu2011chern,fang2012multi,yang2014classification,gao2016classification,liu2017predicted,yu2018nonsymmorphic,wu2020higher,wu2021higher,footnote}. 
Particularly, in two-dimensional (2D) systems, the existence of quadratic nodal points (QNPs), around which the bands disperse quadratically, was noted in several material examples, such as bilayer graphene~\cite{neto2009electronic,mccann2006landau}, blue phosphorene oxide~\cite{zhu2016blue}, monolayer Mg$_2$C~\cite{wang2018monolayer}, monolayer Na$_2$O and K$_2$O~\cite{hua2020tunable}, and 2D Si/Bi heterostructure~\cite{zhao2022two}. It is important to note that: (1) so far, QNP has been found only in nonmagnetic materials; (2) the reported QNPs are stable only in the absence of SOC. Very recently, a systematic classification work indicated the possibility to have a QNP in 2D magnetic systems~\cite{zhang2023encyclopedia}, but no example material was identified. Therefore, it is an interesting and urgent task to find a concrete magnetic QNP material.

In this work, we accomplish this task and identify magnetic QNP in a 2D kagome metal organic framework (MOF) material
Cr$_{3}$(HAB)$_{2}$  (HAB represents the organic ligand hexaaminobenzene)
with sizable SOC. MOFs are a huge class of crystalline materials consisting of metal ions or clusters connected by organic ligands~\cite{li1999design,furukawa2013chemistry}. A big advantage of these materials is their high tunability: current techniques can enable synthesis of various MOFs with designed composition and structure~\cite{farha2010rational}. Magnetism can also be readily introduced into MOFs by incorporating magnetic ions (such as $3d$ elements) \cite{pedersen2018formation,dong2018coronene}. The Cr$_{3}$(HAB)$_{2}$ MOF proposed here is motivated by the recent experiments which successfully synthesized a family of 2D MOFs, including Ni$_{3}$(HAB)$_{2}$ \cite{lahiri2017hexaaminobenzene} and Cr$_{3}$(HITP)$_{2}$ \cite{zhong2023synthesizing}, with the same type of kagome structure.
Here, by first-principles calculations, we demonstrate the stability of 2D monolayer Cr$_{3}$(HAB)$_{2}$. Our spin polarized calculation shows that its ground state is a noncollinear antiferromagnet (AFM) with an in-plane $120^\circ$ configuration.
In this AFM state, the conduction and valence bands form a protected QNP at $\Gamma$ point of BZ. Moreover, this QNP is robust under SOC. We analyze the symmetry protection of this magnetic QNP, and construct effective $k\cdot p$ and lattice models to capture its key features. We show that the magnetic QNP can be the only Fermi point of the system, without additional states at Fermi level.  The optical absorbance of such a system can exhibit a quantized plateau of $\pi\alpha/2$ at low frequencies, where $\alpha$ is the fine structure constant. In addition, as for monolayer Cr$_{3}$(HAB)$_{2}$, besides QNP, there also exist six linear nodal points. Topological edge states and effects of electronic correlations are discussed. Our work reveals the first concrete material which hosts QNP under magnetism and SOC.
It will help to clarifies properties of QNPs, to explore its interplay with magnetic ordering, and to facilitate subsequent experimental studies.

\section{Computation Method}
Our first-principles calculations are based on the density functional theory (DFT) and are performed by using Vienna ab-initio Simulation Package \cite{Kresse1993Ab,Kresse1996}. Projector augmented wave method is used to treat ionic potentials \cite{PAW1994}. Generalized gradient approximation with the Perdew-Burke-Ernzerhof (PBE) parametrization is adopted to model
exchange-correlation functional \cite{Perdew1996}. To account for the possible correlation effects of Cr-3$d$ electrons, we employ the GGA$+U$ method \cite{dudarev1998electron}.  $U$ values ranging from 0 to 3 eV have been tested, and we find the qualitative results regarding magnetic ordering and QNP are not affected. Therefore, we will focus on the $U=0$ result in the following.
The energy cutoff is set to 520 eV. The 2D BZ is sampled with a $\Gamma$-centered grid with a size of $5\times 5\times 1$. The crystal structure is fully relaxed with energy and force convergence criteria being set to 10$^{-7}$ eV and 10$^{-3}$ eV/\AA, respectively. A vacuum region of 20 Å thickness is added to suppress interactions between periodic images of the 2D layer. The edge spectrum is studied by using the WannierTools Package \cite{Wu2018}, with the ab-initio tight-binding Hamiltonian constructed by the WANNIER90 package \cite{Mostofi2008}. The effective models are constructed with the help of MagneticKP \cite{zhang2023magnetickp} and MagneticTB \cite{zhang2022magnetictb} packages. The irreducible corepresentations of states are analyzed by using the IRVSP code~\cite{gao2021irvsp}.

\section{Crystal structure and magnetism}

\begin{figure}
	\includegraphics[width=8.6cm]{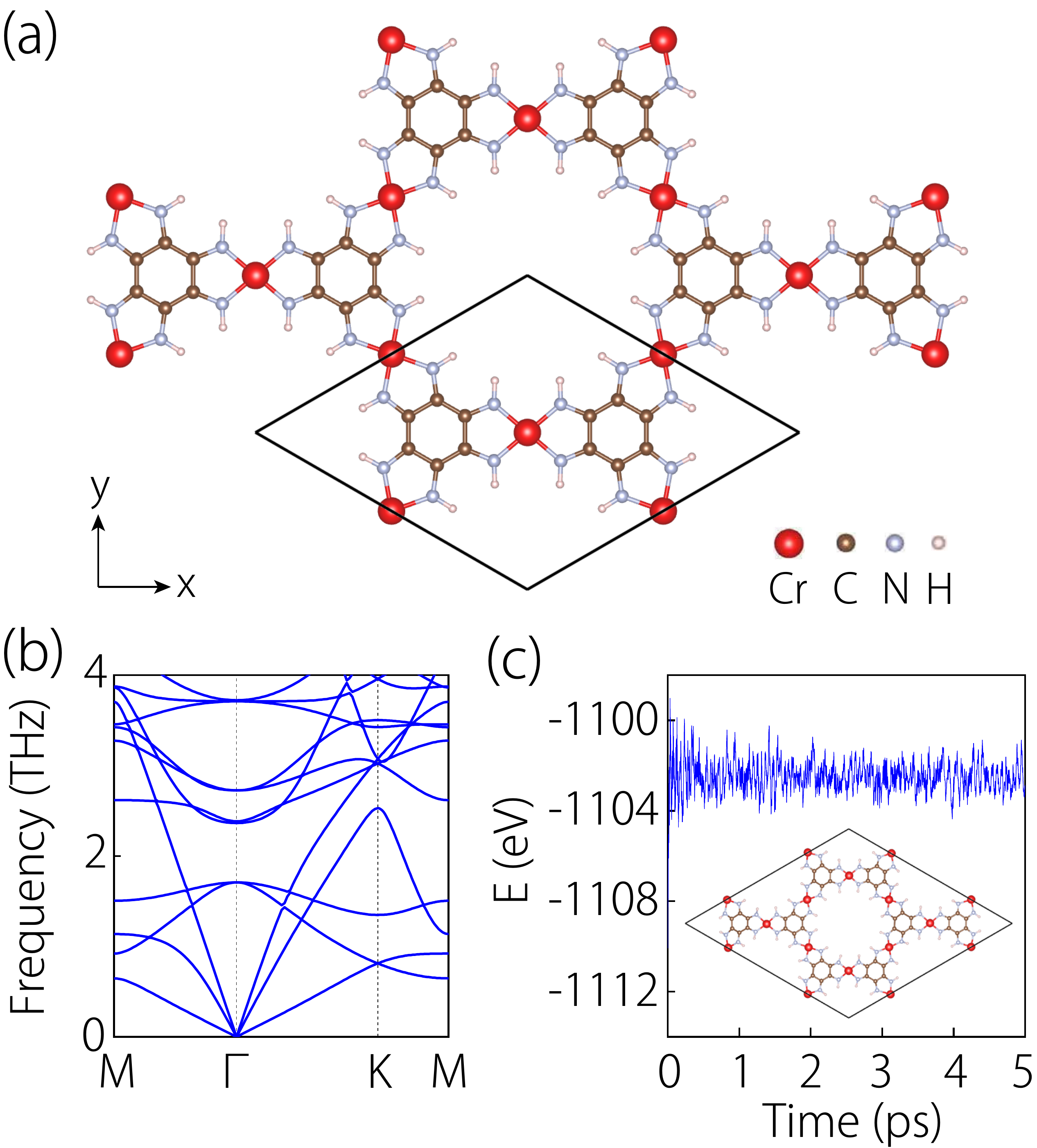}
	\caption{  (a) Top view of the lattice structure of 2D MOF Cr$_{3}$(HAB)$_{2}$. The black box indicates the unit cell. (b) Calculated phonon spectrum, demonstrating the dynamical stability of the structure. (c) Energy versus time in the molecular dynamics simulation for temperature of 400 K. The inset shows the snapshot of the structure after 5 ps simulation period.
		\label{fig1}}
\end{figure}

The crystal structure of monolayer Cr$_{3}$(HAB)$_{2}$ is illustrated in Fig.~\ref{fig1}(a). It has a 2D
hexagonal lattice structure, in which the metal ions (i.e., Cr$^{2+}$) form a kagome sublattice.
Several transition metal MOFs with the same type of structure have been successfully synthesized in experiment, such as
Ni$_{3}$(HAB)$_{2}$, Cu$_{3}$(HAB)$_{2}$, and Co$_{3}$(HAB)$_{2}$ \cite{lahiri2017hexaaminobenzene,hinckley2020air}.  From our calculation, the optimized lattice constant is 13.70 \AA\ [see Fig.~\ref{fig1}(a)]. The monolayer crystal is completely flat with single-atom thickness. The crystal lattice has space group $P6/mmm$ (No.~191), and its point group is $D_{6h}$. In Fig.~\ref{fig1}(a), we also indicate the unit cell, which contains three Cr sites forming an equilateral triangle. The HAB ligands bridge the Cr sites through forming Cr-N bonds.

To investigate the stability of Cr$_{3}$(HAB)$_{2}$, we have calculated its phonon spectrum. The result is plotted in Fig.~\ref{fig1}(b). One observes that there is no imaginary frequency mode, which verifies the dynamic stability of monolayer Cr$_{3}$(HAB)$_{2}$. We also perform
ab-initio molecular dynamics simulations to check its thermal stability.
The simulated time duration is 5 ps (5000 steps), and the ambient temperature is set to 400 K. The obtained free energy evolution and the snapshot for the final structure are displayed in Fig.~\ref{fig1}(c). The result indicates that monolayer Cr$_{3}$(HAB)$_{2}$ has good thermal stability and maintains its overall structure at 400 K.


\begin{figure*}
	\includegraphics[width=17.5cm]{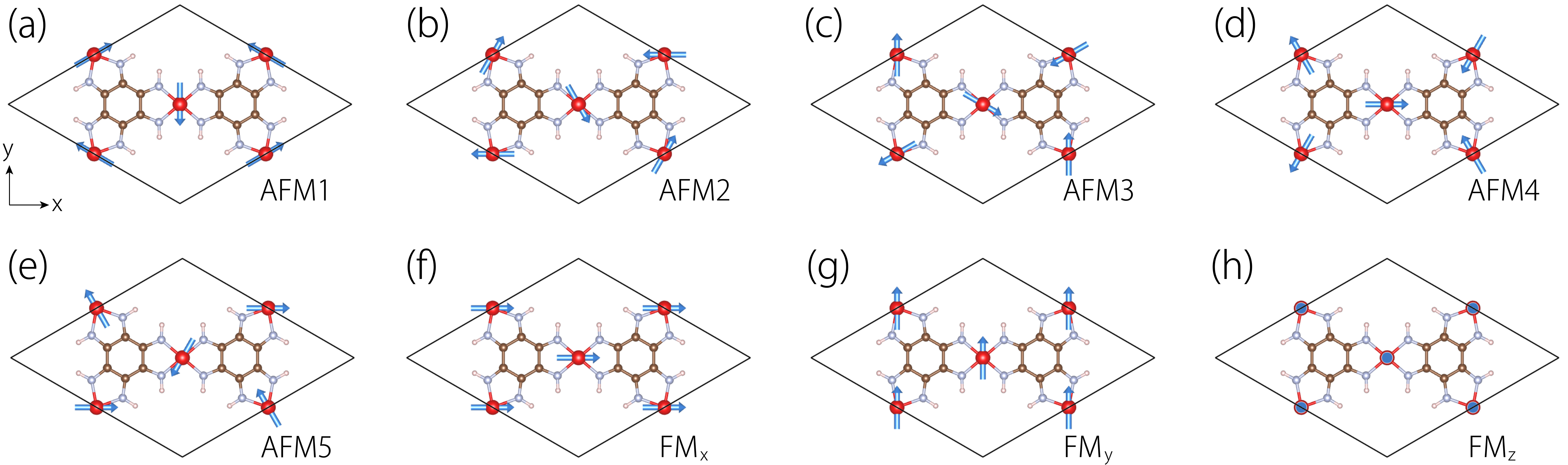}
	\caption{ Magnetic configurations of Cr$_{3}$(HAB)$_{2}$ considered in the calculation. (a) is the noncollinear AFM configuration dubbed AFM1, which is found to be the ground state configuration. (b), (c) and (d) are obtained from (a) by rotation all local spins by 30$^\circ$, 60$^\circ$ and 90$^\circ$, respectively. (e) is the configuration with opposite spin chirality compared to (d). (f-h) show the three FM configurations.
		\label{fig2}}

\end{figure*}
\begin{table*}
	\caption{\label{table1} Comparison of total energies (per unit cell) for different magnetic configurations. The values are in unit of meV, with reference to the energy of AFM1 configuration. Here, we consider Hubbard $U$ values from 0 to 3 eV.}
	\begin{ruledtabular}
		\begin{tabular}{ccccccccc}		
			\quad U &\qquad AFM1 &\qquad AFM2 &\qquad AFM3 & \qquad AFM4  & \qquad AFM5 & \qquad FM$_{x}$ & \qquad FM$_{y}$ & \qquad FM$_{z}$ \\
			\hline 	
			\quad 0 &\qquad 0 &\qquad 0.105 & \qquad 0.319 & \qquad 0.426 & \qquad
                3.473 & \qquad 224.352 & \qquad 224.357 & \qquad 223.251 \\
                \quad 1 &\qquad 0 &\qquad 0.130 & \qquad 0.391 & \qquad 0.523 & \qquad 4.172 & \qquad 246.198 & \qquad 246.206 & \qquad 244.806 \\
                \quad 2 &\qquad 0 &\qquad 0.160 & \qquad 0.473 & \qquad 0.627 & \qquad 4.477 & \qquad 260.714 & \qquad 260.707 & \qquad 259.469 \\
                \quad 3 &\qquad 0 &\qquad 0.168 & \qquad 0.504 & \qquad 0.672 & \qquad 4.278 & \qquad 244.254 & \qquad 244.248 & \qquad 243.076 \\
		\end{tabular}
	\end{ruledtabular}
\end{table*}

Next, we investigate the magnetic properties of monolayer Cr$_{3}$(HAB)$_{2}$. Cr typically carries magnetic moment due its partially filled $3d$ shell. This is also the case in Cr$_{3}$(HAB)$_{2}$. To find its ground state magnetic ordering, we
compare calculated energies for various magnetic configurations that are typically tested on a kagome lattice \cite{essafi2017generic}, as schematically shown in Fig.~\ref{fig2}. Besides ferromagnetic (FM) configuration, due to geometric frustration of kagome lattice, the typical AFM configurations are of noncollinear type, where the neighboring local spins make $120^\circ$ angle with each other. SOC is fully considered in our calculation to capture magnetic anisotropy. The calculation result is shown in Table \ref{table1}. One observes that for all the $U$ values considered, the noncollinear AFM1 configuration [Fig.~\ref{fig2}(a)] always has the lowest energy. Thus, the ground state of monolayer Cr$_{3}$(HAB)$_{2}$ has magnetic ordering of AFM1 type. Our calculation also shows that the magnetic moment is mainly on the Cr site, as expected, and has a magnitude $\sim 2.8 \mu_B$. With the AFM1 configuration, the magnetic space group of the system is $P6^{\prime} / m^{\prime} m m^{\prime}$, which is a type-III magnetic space group.


\begin{figure}
	\includegraphics[width=8.6cm]{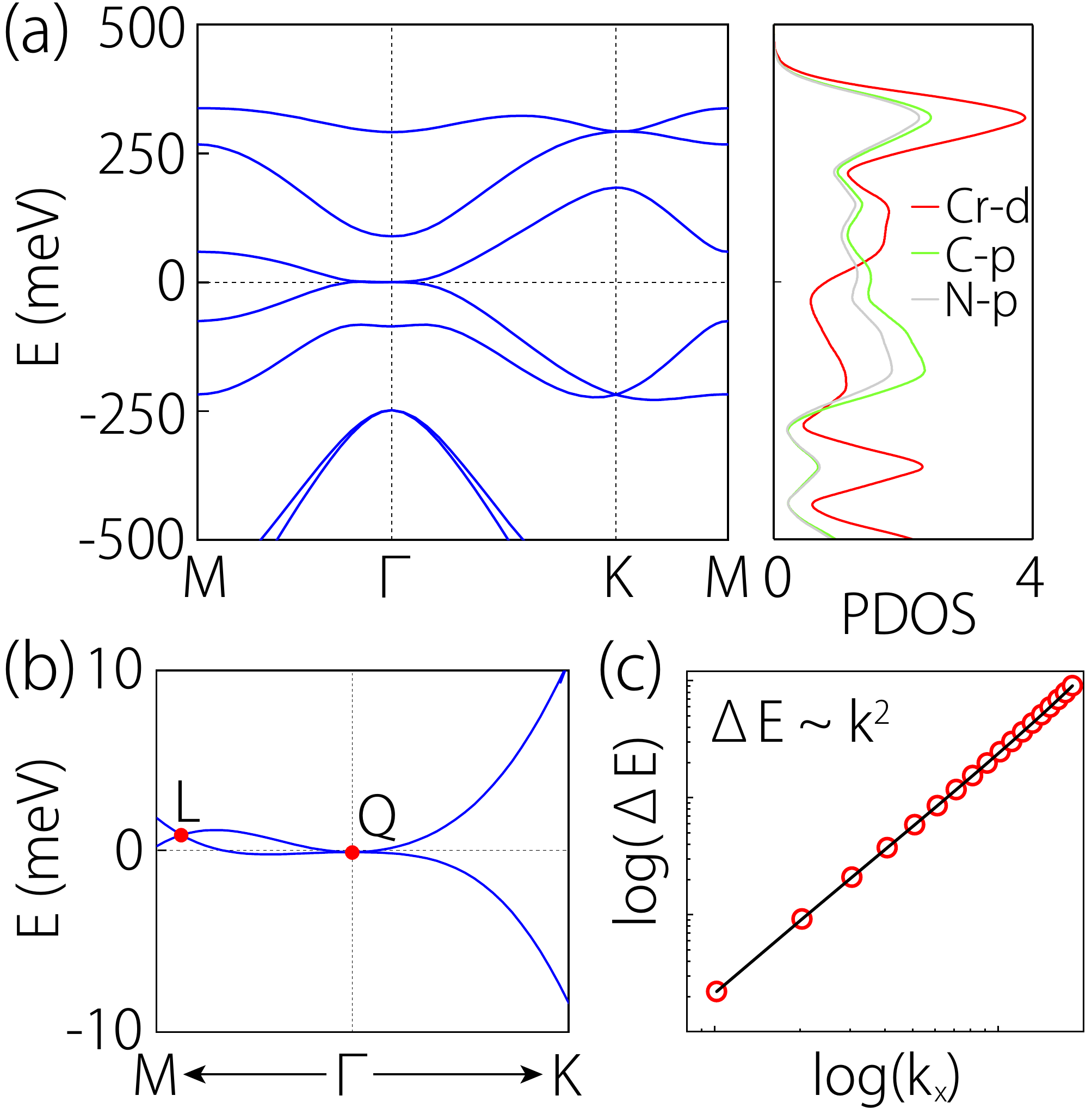}
	\caption{  (a) Calculated band structure and projected density of states of Cr$_{3}$(HAB)$_{2}$ for the ground state (AFM1) configuration. (b) Enlarged view of band structure around QNP (point $Q$).  $L$ indicates a linear nodal point. (c) Band splitting versus $k$ around QNP in log-log plot, which has a slope of 2, confirming the quadratic dispersion.
		\label{fig3}}
\end{figure}

\section{magnetic QNP}
As mentioned above, the previously reported QNPs all appear in nonmagnetic materials and in the absence of SOC. In this section, we show that a QNP appears in AFM Cr$_{3}$(HAB)$_{2}$ in the presence of SOC.

In Fig.~\ref{fig3}(a), we plot the calculated electronic band structure (with the ground state AFM1 configuration) along with the projected density of states (PDOS). One observes that the low-energy bands are mainly from Cr-3$d$, C-$p$ and N-$p$ orbits, and the system shows a typical semimetal character. By zooming in the band structure around $\Gamma$ and Fermi level [see Fig.~\ref{fig3}(b)], we find that the conduction and valence bands in fact touch each other at $\Gamma$. We will label this degeneracy point as $Q$. (There is also a crossing point $L$ marked in the figure, which we shall discuss in a while.) The bands in Fig.~\ref{fig3}(a) are non-degenerate, so point $Q$ is twofold degenerate. Importantly, the dispersion for both conduction and valence bands around $Q$ appears to be of nonlinear type, which indicates the possible existence of a higher-order nodal point.

To investigate the order of dispersion, in Fig.~\ref{fig3}(c), we plot the band splitting, i.e., the energy difference between conduction and valence bands, versus momentum $k$ near $Q$ point. The result clearly shows a quadratic scaling, suggesting that $Q$ is a QNP. To pin down its QNP character, we construct the $k\cdot p$ model expanded at $Q$ based on  symmetry.
The little co-group at $\Gamma$ is $6^{\prime} / m^{\prime} m m^{\prime}$, which is generated by inversion $P$
, two-fold rotation $C_{2y}$ 
and magnetic operation $C_{6z}T$, where $T$ is the time reversal operator
. The twofold band degeneracy at $Q$ corresponds to the $\Gamma_{4}^{-}$ irreducible corepresentation of $6^{\prime} / m^{\prime} m m^{\prime}$.
In the basis of the two degenerate states at $Q$, the above three generators take the following matrix representations:
\begin{equation}
P=-\sigma_{0}, \quad
C_{2y}=i\sigma_{z},\quad
C_{6z}T=-e^{i\pi\sigma_y/3},
\end{equation}
where the Pauli matrices $\sigma_i$'s are acting on the space spanned by the two basis states.
Constrained by these symmetries, we obtain the following $k \cdot p$ effective Hamiltonian expanded to $k^2$ order at $Q$:
\begin{equation}\label{eq1}
\mathcal{H}_{Q} = a_1k^{2}\sigma_{0} + a_{2}(k_{y}^{2} - k_{x}^{2})\sigma_{z} + 2a_{2}k_{x} k_{y} \sigma_{x},
\end{equation}
where  $a_{1}$ and $a_{2}$ are two real model parameters, and the energy is measured from $Q$.
One can see that the $k$-linear terms are forbidden by symmetry, and the leading order is of $k^2$. This confirms that the point $Q$ here is a QNP.

It usually requires high symmetry to stabilize band degeneracies with higher-order dispersions \cite{wu2021higher,yu2019quadratic}. In this regard, the noncollinear magnetic ordering is in fact essential for the emergence of QNP here, since it to a large extent preserves the symmetry of the lattice. SOC is also essential here. The QNP will disappear if SOC is removed. In fact, the systematic classification in Ref.~\cite{zhang2023encyclopedia} showed that QNP cannot exist in magnetic systems without SOC.

Our next question is: Can such magnetic QNP be the only Fermi point of a system? In other words, is it possible to have a QNP semimetal in which the Fermi level cut through only the QNP but no other electronic states? This is not apparent from the band structure of Cr$_{3}$(HAB)$_{2}$ in Fig.~\ref{fig3}(a), since there clearly exists  another crossing $L$.

\begin{figure}
	\includegraphics[width=8.6cm]{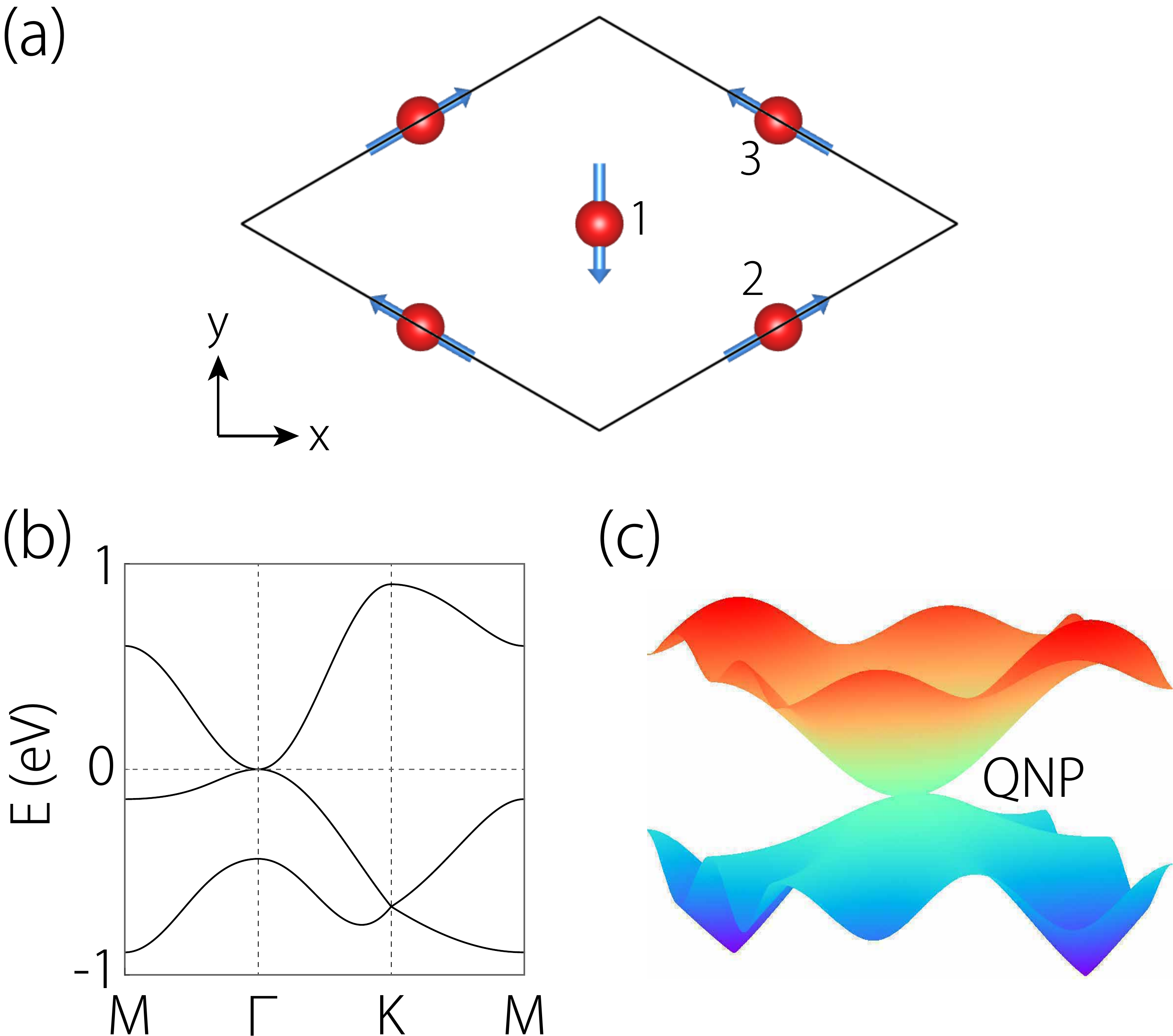}
	\caption{(a) Schematic of the minimal lattice model. In a unit cell, we put three spin polarized basis. See the main text for description. (b) Band structure of the lattice model with parameter values $\varepsilon_0 = -0.15$, $t_1 = 0.155$ and $t_2 = -0.23$. (c) shows the band dispersion around the QNP at $\Gamma$ point in (b).
		\label{fig4}}
\end{figure}

To address this question and also to facilitate future studies of magnetic QNPs, we attempt to construct a minimal lattice model here. We take a 2D kagome lattice, with three active sites in a unit cell, as shown in Fig.~\ref{fig4}(a). On each site, we put an $s$-like orbital.
To incorporate the AFM1 ordering, we assume each basis orbital is spin polarized and has polarization along the local moment direction. Labelling the three basis orbitals as $|i\rangle$ $(i=1,2,3)$ [see Fig.~\ref{fig4}(a)], we have
\begin{equation}
\begin{split}
&\ket{1}=C_{4x}\ket{\uparrow}=e^{\frac{i\pi\sigma_x}{4}}\ket{\uparrow},\\
&\ket{2}=C_{3z}\ket{1}=e^{\frac{i\pi\sigma_z}{3}}\ket{1},\\
&\ket{3}=C_{3z}\ket{2}=e^{\frac{i\pi\sigma_z}{3}}\ket{2},\\
\end{split}
\end{equation}
where $\ket{\uparrow}$ is the usual spin up state polarized along $z$ direction. In this basis, the symmetry generators
of the magnetic space group take the following representations
%
%
%
 \begin{equation}
 	\begin{split}
 		P=\begin{pmatrix}
 			1&0&0\\
 			0&1&0\\
 			0&0&1\\
 		\end{pmatrix},\qquad
 		C_{2y}=\begin{pmatrix}
 			i&0&0\\
 			0&0&-i\\
 			0&-i&0\\
 		\end{pmatrix},\\
 		C_{6z}T=\begin{pmatrix}
 			0&0&i\\
 			-i&0&0\\
 			0&-i&0\\
 		\end{pmatrix}.
 	\end{split}
 \end{equation}
Constrained by these symmetries, we obtain the following lattice model including hopping terms up to second neighbor:
\begin{equation}\label{latt}
    \begin{aligned}
    \mathcal{H}_\text{latt} = \ \varepsilon_0 &+\Big(t_1 \cos\frac{\sqrt{3}k_x-k_y}{4}+t_2 \cos\frac{\sqrt{3}k_x+3k_y}{4}\Big)\Lambda_{1}\\
    &+\Big(t_1 \cos\frac{\sqrt{3}k_x+k_y}{4}+t_2 \cos\frac{\sqrt{3}k_x-3k_y}{4}\Big)\Lambda_{4}\\
    &+\Big(t_1 \cos\frac{k_y}{2}+t_2 \cos\frac{\sqrt{3}k_x}{2}\Big)\Lambda_{6},
    \end{aligned}
\end{equation}
where $\varepsilon_0$ is an overall energy shift, $t_{1}$ and $t_{2}$ represent the first and the second neighbor hopping, respectively, $k$ is in unit of $a^{-1}$ ($a$ is the lattice constant of the lattice model), and $\Lambda_i$'s are the $3 \times 3$ Gell-Mann matrices with
\begin{equation}
\begin{aligned}
&\Lambda_1=\left(\begin{array}{ccc}
0 & 1 & 0 \\
1 & 0 & 0 \\
0 & 0 & 0
\end{array}\right),\quad
\Lambda_4=\left(\begin{array}{ccc}
0 & 0 & 1 \\
0 & 0 & 0 \\
1 & 0 & 0
\end{array}\right),\quad\\
&\Lambda_6=\left(\begin{array}{ccc}
0 & 0 & 0 \\
0 & 0 & 1 \\
0 & 1 & 0
\end{array}\right).
\end{aligned}
\end{equation}

A typical band structure of our minimal lattice model $\mathcal{H}_\text{latt}$ is shown in Fig.~\ref{fig4}(b).
It indeed reproduces the magnetic QNP and the qualitative features of the low-energy bands (the conduction band and the top two valence bands) in monolayer Cr$_{3}$(HAB)$_{2}$. Notably, the result shows that it is indeed possible to have QNP as the only Fermi point, without any additional states at Fermi energy, as displayed in Fig.~\ref{fig4}(b). Thus, one can have a well defined magnetic QNP semimetal state.

\section{linear nodal points}

\begin{figure}
	\includegraphics[width=8.6cm]{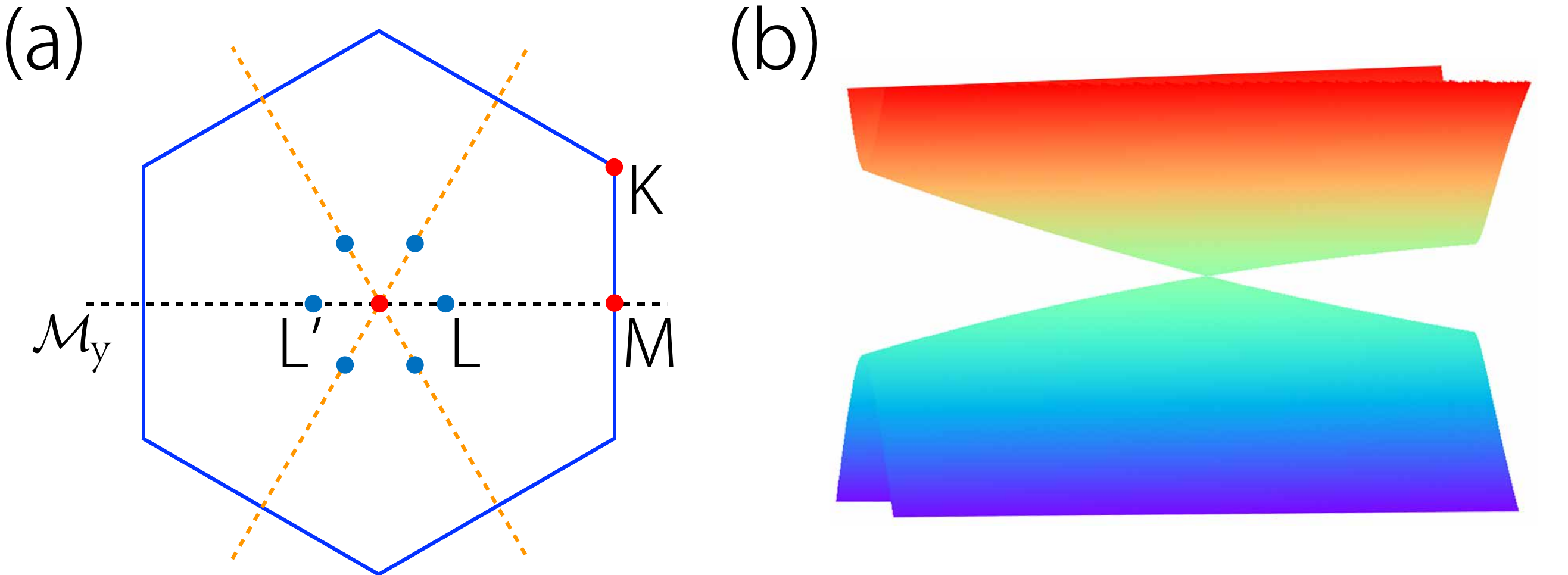}
	\caption{ (a) Distribution of the six linear nodal points (indicated by the blue dots) in Brillouin zone.
$L$ and $L'$ are located on the mirror line ($M_y$). The other four points are related to $L$ and $L'$
by $MT$ or $M'T$ operations.
 (b) Band dispersion around $L$ point, demonstrating the linear type dispersion.
	\label{fig5}}
\end{figure}

Now, let us turn to the crossing point $L$ on $\Gamma$-$M$ path. Due to $C_{6z}T$ symmetry, there are actually six such linear nodal points in BZ, as illustrated in Fig.~\ref{fig5}(a). The linear band dispersion around $L$ point is shown in Fig.~\ref{fig5}(b). We can also construct a $k\cdot p$ model for its description. The magnetic little co-group on $\Gamma$-$M$ path ($k_x$ axis) is $m^{\prime} m 2^{\prime}$, generated by vertical mirror $M_{y}$ and magnetic symmetry $C_{2z}T$. The two crossing bands on this path correspond to the one-dimensional irreducible corepresentations $\Sigma_{3}$ and $\Sigma_{4}$. Following similar approach as in the last section, we obtain the following effective model
expanded at $L$ point 
\begin{eqnarray}\label{eq6}
\mathcal{H}_L = b_1q_{x}\sigma_{0} + b_2q_{x}\sigma_{z} + b_3q_{y}\sigma_{x},
\end{eqnarray}
where the energy and the momentum $\bm q=(q_{x},q_{y})$ are measured from $L$. This is a typical tilted linear Weyl point model, with the $b_1$ term describing the tilt along the mirror line.

It is worth noting that point $L$ enjoys a double protection, by $C_{2z}T$ and by the mirror line.
The mirror protection is straightforward, as the two crossing bands have opposite mirror eigenvalues, and this constrains the nodal points to be located on $\Gamma$-$M$.
Meanwhile, the $C_{2z}T$ symmetry alone can also protect the points. This is because as an anti-unitary operator, $(C_{2z}T)^2=1$, and this ensures quantization of Berry phase (in unit of $\pi$) along any closed path. Each linear point here features a $\pi$ Berry phase, so it cannot be gapped by itself  under perturbations that respect
$C_{2z}T$ (only pair annihilation is allowed). One can imagine that if some perturbation breaks the mirror lines but preserves  $C_{2z}T$, then these linear Weyl point can still exist and become fully unpinned in BZ (i.e., they may move to generic $k$ points)\cite{lu2016multiple,lu2022realization}.

\section{Edge states}

\begin{figure}
	\includegraphics[width=8.6cm]{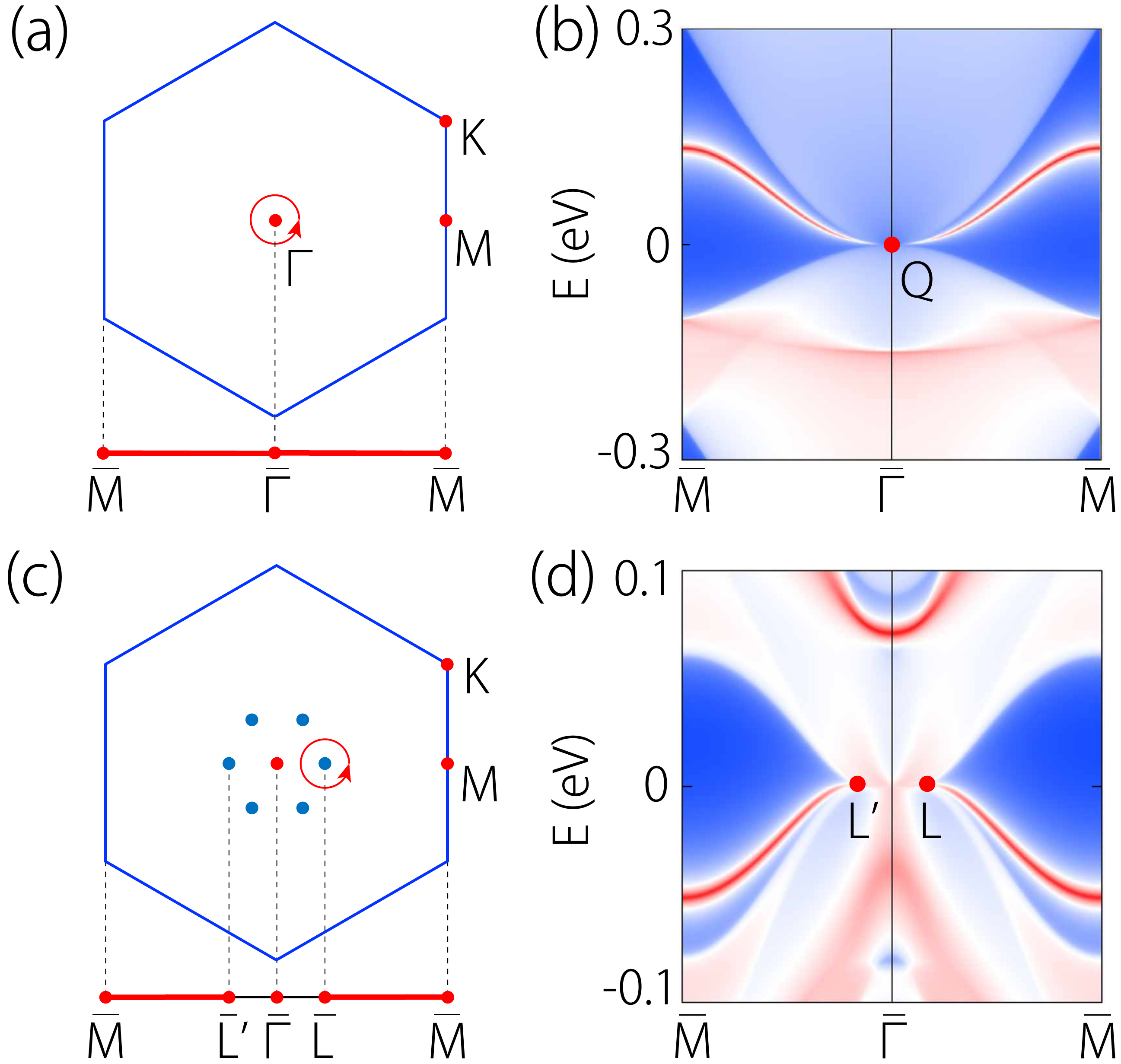}
	\caption{  (a) Bulk and edge Brillouin zones of the minimal lattice model in Eq.~(\ref{latt}. QNP is marked by red dot at $\Gamma$ point. (b) Edge spectrum for the minimal lattice model. One observes edge bands across the whole edge Brillouin zone, resulted from the $\pi$ Zak phase in the bulk. (c) For 2D Cr$_{3}$(HAB)$_{2}$, besides QNP, there also exist six linear nodal points. The Zak phase is $0$ in the central region between $\bar{L}$ and $\bar{L}'$ and is $\pi$ outside. This leads to the edge spectrum shown in (d).
		\label{fig6}}
\end{figure}

In this section, we investigate the edge spectrum of a magnetic QNP semimetal state. We first consider the
minimal lattice model in Eq.~(\ref{latt}). In Fig.~\ref{fig6}(b), we plot the calculated edge spectrum for an edge along the $x$ direction. One observes an edge band, which appears throughout the whole edge BZ and connects to the edge projection of QNP.
The existence of this edge band is not directly related to QNP. Instead, it corresponds to the quantized $\pi$ Berry phase (Zak phase) for the straight path at each fixed $k_x$ traversing the bulk BZ. (Again, the quantization is due to $C_{2z}T$.) Nevertheless, it should be pointed out that it is because QNP features a $2\pi$ (or zero) Berry phase that edge states can appear over the whole edge BZ, as illustrated in Fig.~\ref{fig6}(a). Similar boundary states that occupy the whole surface BZ for a 3D system were previously reported for states with higher-order nodal lines \cite{yu2019quadratic,zhang2021magnetic}. In Ref.~\cite{zhang2021magnetic}, such boundary states are called torus surface states, since the surface BZ has the topology of a torus.

Next, we consider the edge spectrum of Cr$_{3}$(HAB)$_{2}$. The result is plotted in Fig.~\ref{fig6}(d). At first glance, the qualitative features of Fig.~\ref{fig6}(d) seem to be unchanged from Fig.~\ref{fig6}(b). However, there is difference due to the presence of linear nodal points.
As indicated in Fig.~\ref{fig6}(c), edge projections of the two outer linear points, marked as $\bar{L}$ and $\bar{L}'$, divide the edge BZ into two regions: the inner region including $\bar{\Gamma}$ point and the outer region including  $\bar{M}$ point.
Since each linear point has a $\pi$ Berry phase, the Zak phases for the two regions must differ by $\pi$. In the current case, $\pi$ Zak phase appears for the outer region, which dictates the edge states there. This physics is similar to what happens in graphene on the zigzag edge \cite{neto2009electronic,yao2009edge}. Nevertheless, one should note that for graphene, the edge states are spin degenerate (if we do not consider possible edge magnetism due to correlations), whereas for Cr$_{3}$(HAB)$_{2}$, the edge band is spin polarized.


\section{Discussion and Conclusion}

A QNP semimetal state can possess interesting physical properties. For example, it may exhibit a quantized plateau in low-frequency optical absorbance. Optical absorbance $A(\omega)=W_a/W_i$ is the ratio between the absorbed energy flux $W_a$ and the incident flux $W_i$ of light \cite{nair2008fine}. At low frequencies where the band dispersion is dominated by quadratic terms,
assuming the optical transition involves only the two low-energy bands as described by $\mathcal{H}_Q$ in Eq.~(\ref{eq1}),
then $A(\omega)$ can be readily evaluated from Fermi's golden rule. In this case, one finds a quantized value
$A(\omega)=\pi\alpha/2\approx 1.15\%$, where $\alpha=e^2/\hbar c \approx 1/137$ is the fine structure constant.
Such quantized absorbance was previously reported in graphene \cite{nair2008fine}, blue phosphorene oxide \cite{zhu2016blue}, and other 2D nodal-point semimetals \cite{wu2021higher}
in the absence of SOC. For graphene and blue phosphorene oxide, the quantized value is $\pi\alpha$. The spinful QNP here is like half of the spin-degenerate QNP in blue phosphorene oxide, which explains the difference of $1/2$ factor.

For the QNP in monolayer Cr$_{3}$(HAB)$_{2}$, we have mentioned that it is quite robust against the variation of $U$ values \cite{SupplementaryMaterial}.
We have tested $U$ values ranging from 0 to 3 eV, and the QNP is always maintained, just its energy slightly shifted down with increasing $U$. Meanwhile, when $U$ increases, the linear Weyl points will shift outward along the $\Gamma$-$M$ path, and at $U>2$ eV, two points will collide into each other and then annihilate (see Supplemental Material \cite{SupplementaryMaterial}).

In previous studies \cite{wu2021higher}, it was shown that for spinless QNP (i.e., in the absence of SOC) in nonmagnetic materials, the minimal symmetry requirement is $n$-fold rotation $C_{nz}$ with $n>2$ and time reversal symmetry $T$. As for magnetic QNP,
considering type-III magnetic space groups, we find the minimal symmetry group is $P \bar{6}^{\prime}$, which can be generated by $C_{3z}$ and $M_z T$ ($M_z$ is the horizontal mirror plane).  One notes that the symmetry of Cr$_{3}$(HAB)$_{2}$ with AFM1 ordering is larger than this. Hence, some symmetry (e.g., $C_{2y}$) of Cr$_{3}$(HAB)$_{2}$
may be broken without affecting the stability of QNP. For example, we may consider rotating the local spins all by the same angle $\theta$ [see for example Fig.~\ref{fig2}(b)]. Clearly, this preserves both $C_{3z}$ and $M_z T$, thus the QNP should still be maintained.
In Supplemental Material \cite{SupplementaryMaterial}, we present a calculation result, which confirms this point.

Finally, we discuss some experimental aspects. As we have mentioned, although Cr$_{3}$(HAB)$_{2}$ has not been realized yet, the chance to synthesize it in the near future is quite promising. Experimental techniques have been developed to fabricate
a rich variety of MOFs with designed compositions \cite{li1999design,furukawa2013chemistry}.
The closely related 2D MOFs, especially the transition metal MOFs $M_{3}$(HAB)$_{2}$ with $M=$ Ni, Cu, Co \cite{hinckley2020air}, and also Cr$_{3}$(HITP)$_{2}$ \cite{zhong2023synthesizing}, have already been successfully synthesized. Our work here will offer an impetus for experiment on this material. Once achieved, the QNP in the band structure can be imaged by angle resolved photoemission spectroscopy (ARPES) \cite{lv2021experimental,sobota2021angle}. The topological edge states can be probed by ARPES or scanning tunneling spectroscopy (STS) \cite{zheng2016atomic}. The spin polarization of bulk bands and edge states may also be probed by STS by using magnetic tips.


In conclusion, we propose the first material realization of a magnetic spinful QNP in the 2D MOF Cr$_3$(HAB)$_2$.
Interestingly, such a higher-order nodal point occurs in presence of a noncollinear magnetic ordering and sizable SOC, which is quite unusual. The constructed effective $k\cdot p$ and lattice models capture the essential features of magnetic QNP, and they form a good starting point for subsequent studies on QNP fermions. Features of coexisting linear Weyl points, topological edge states, quantized optical absorbance, and robustness of these nodal points are discussed. Our findings will facilitate the exploration of novel 2D emergent fermions and their interplay with magnetism.

\begin{acknowledgments}
The authors thank D. L. Deng for valuable discussions. This work is supported by the NSF of China (Grant No. 12004028 and No. 12174018), the Shandong Provincial Natural Science Foundation (Grant No. ZR2023QA012), the Special Funding in the Project of Qilu Young Scholar Program of Shandong University and Singapore NRF CRP (CRP22-2019-0061). We acknowledge computational support from the Texas Advanced Computing Center and the National Supercomputing Centre Singapore.
\end{acknowledgments}

\bibliography{Magnetic_QWP_v1009.bib}

\bibliographystyle{apsrev4-2}

\end{document}